\documentstyle[12pt,aasms4]{article}

\begin{document}
\oddsidemargin=15mm
\onecolumn

\title{
A High Intrinsic Peculiarity Rate Among Type Ia Supernovae
}

\author{Weidong Li, Alexei V. Filippenko, Richard R. Treffers }
\affil{Department of Astronomy, University of California, Berkeley, CA 94720-3411}

\author{Adam G. Riess}
\affil{Space Telescope Science Institute, 3700 San Martin Drive, Baltimore, MD 21218}

\centerline{and}

\author{ Jingyao Hu and Yulei Qiu}
\affil{Beijing Astronomical Observatory, Chinese Academy of Sciences, Beijing 100080, P. R. C.}

\newpage

\begin{abstract}
 
We have compiled a sample of 45 Type Ia supernovae (SNe Ia) discovered
by the Lick Observatory Supernova Search (LOSS) and
the Beijing Astronomical Observatory Supernova Survey (BAOSS),
and determined the rate of spectroscopically peculiar 
SNe Ia (i.e., SN 1986G-like, SN 1991bg-like, and SN 1991T-like objects) 
and the luminosity function of SNe Ia.
Because of the nature of the two surveys (distance-limited with 
small baselines and deep limiting magnitudes), 
nearly all SNe Ia have been
discovered in the sample galaxies of LOSS and BAOSS; thus, the observed
peculiarity rate and luminosity function of SNe Ia are intrinsic.
We find that 36$\pm$9\% of nearby SNe Ia are peculiar; specifically, the luminosity
function of SNe Ia consists of 20\% SN 1991T-like, 64\%  normal, and
16\% SN 1991bg-like objects.
 We have compared our results to those
found by earlier studies, and to those found at high redshift. The
apparent dearth of SN 1991T-like objects
at high redshift may be due to extinction, and especially to the difficulty of 
recognizing them from spectra obtained past maximum
brightness or from spectra with low signal-to-noise ratios.
Implications of the high peculiarity rate for the progenitor systems of 
SNe Ia are also briefly discussed.

\end{abstract}


\section{ Introduction}

Supernovae (SNe) are classified as Type Ia on the basis of distinguishing
features in their optical spectra (e.g., Harkness \& Wheeler 1990). During
the early photospheric phase, for example, SN Ia spectra lack conspicuous
lines of hydrogen and contain a strong absorption feature at about 6150 \AA\,
due to blueshifted Si II $\lambda$6355.
The inferred expansion velocity of the ejecta is more than 10,000 km s$^{-1}$, 
implying a violent explosion.

A striking characteristic of SNe Ia is that they display a remarkable 
degree of similarity, more than any other subclass of SNe, and their
spectral evolution follows a reproducible pattern (e.g., Kirshner et al. 1973; Branch \& Tammann 1992; Wells et al. 1994; Filippenko 1997). 
The light curves
of SNe Ia are also impressively homogeneous (e.g., Leibundgut et al. 1992;
Branch \& Tammann 1992). At maximum brightness the color is strongly peaked
near $B - V $ = 0 mag (Patat et al. 1997), and the absolute
magnitude has a much smaller dispersion than in other subclasses 
of SNe. For example, the dispersion of 
the Cal\'an/Tololo Survey is only $\sim$ 0.2 mag when a few 
observationally red SNe Ia are excluded (Hamuy et al. 1996).
Thus, SNe Ia have been considered homogeneous and used as 
``standard candles" (see Branch \& Tammann 1992 for a review).

As summarized by Filippenko (1997) and in 
Section 2, however, the existence of the peculiar SN Ia 1986G and the 
appearance of two very peculiar SNe Ia (1991T and 1991bg) in the same 
year have drawn attention to diversity among SNe Ia.
These variations have raised concerns about the cosmological
utility of SNe Ia. 
{In an attempt to quantify the rate of spectroscopically peculiar 
SNe Ia in the existing observed sample, Branch, Fisher, \&  Nugent
(1993, hereafter BFN) compiled a set of 84 SNe Ia, and subclassified them as
either normal or like one of the known peculiar SNe Ia: SN 1986G, SN 1991T,
or SN 1991bg. They found that their sample is affected by the 
Malmquist bias; nevertheless, about 83\% to 89\% of
their sample is normal.} 
Even among the relatively normal SNe Ia, however, there are 
observed differences in spectra and light curves
(e.g., Filippenko 1997; Hamuy et al. 1996; Riess et al. 1999a).
Fortunately, the variation in luminosity 
is correlated with light-curve shape,
 providing a means to calibrate the peak luminosity of 
 SNe Ia [e.g., the $\Delta m_{15}(B)$ method -- Phillips 1993,
Hamuy et al. 1996, Phillips et al. 1999; the multi-color light-curve shape method --
Riess, Press, \& Kirshner 1996, Riess et al. 1998; the stretch method -- Perlmutter et al. 1997]. Moreover, 
the photometric sequence exhibited  by SNe Ia 
represents a spectroscopic sequence, from overluminous (by $\sim$0.4 mag in $V$)
SN 1991T-like objects to subluminous (by $\sim$2 mag in $V$) 
SN 1991bg-like objects (Nugent et al. 1995).

The intrinsic peculiarity rate of SNe Ia has
implications for our understanding of SNe Ia, especially their progenitor
systems. SNe Ia are thought to come from thermonuclear explosions of 
mass-accreting white dwarfs in  binary systems, when their masses reach
the critical Chandrasekhar mass ($\sim$ 1.4 $M_\odot$; for a review see 
Branch et al. 1995). The exact 
nature of the progenitors, however, is still controversial,
and many scenarios have been proposed. 
If the majority of SNe Ia are normal, we might expect SNe Ia
to come from one class of progenitor system, and the peculiar SNe Ia
to come from some variation of the same system; on the other 
hand, if a significant fraction of SNe Ia are peculiar, 
additional classes of progenitor systems might be warranted.

In the past several years, many SNe Ia have
been discovered in the course of two successful nearby SN surveys --
the Lick Observatory Supernova Search (LOSS; Treffers et al. 1997; Li et al.
2000a; Filippenko et al. 2000) with 
the Katzman Automatic Imaging Telescope (KAIT), and the Beijing
Astronomical Observatory Supernova Survey (BAOSS; Li et al. 1996) with
a 0.6-m telescope. Here
we study the peculiarity rate of SNe Ia from these two surveys.
In Sec. 2 we discuss the sample of SNe Ia used in the statistics,
which consists of 45 SNe Ia discovered by LOSS and BAOSS (up through SN 2000A).
Sec. 3 reports 
the observed and intrinsic peculiarity rate and luminosity function.
The implications of our statistics are discussed in Sec. 4,  and Sec. 5
summarizes our conclusions.

\section{The Sample of Type Ia SNe}

\subsection{ Definition of Peculiar SNe Ia}

As we attempt to compare our results to those of BFN,
we follow their definition of normal and peculiar SNe
Ia. A summary of this definition is given here; see also Filippenko (1997).

A ``normal" SN Ia is defined as one whose optical spectra resemble 
those of SNe 1981B (Branch et al. 1983), 1989B (Barbon et al. 1990),
1992A (Kirshner et al. 1993), and 1972E (Kirshner et al. 1973). 
Normal SNe Ia near maximum light show conspicuous absorption
features near 6150 \AA\, due to Si II and near 3750 \AA\, due to Ca II.
Other absorption features include those of Co II, S II, O I, and
Mg II, as well as additional lines of Si II and Ca II. 
Two weeks after maximum the Si II 6150 \AA\, and the 
Ca II 3750,
8300 \AA\, lines are strong, several Fe II lines develop at the 
blue end of the spectrum, and the absorption near 5700 \AA\, is usually 
attributed to Na I. Blends of forbidden emission lines of iron 
and cobalt dominate the nebular phase, starting about a month
past maximum brightness.

A ``spectroscopically peculiar" SN Ia is defined as one that has 
feature strengths (not just expansion velocities) that differ
significantly from those of
the normal ones at a given phase. Three well-observed examples,
SNe 1991T, 1991bg, and 1986G, serve as prototypes of peculiar 
SNe Ia.

Spectra of SN 1991T (Filippenko et al. 1992a; Phillips et al. 1992; 
Ruiz-Lapuente et al. 1992; Jeffery et al. 1992; Mazzali et al. 1995) 
before maximum brightness are strikingly peculiar
in having unusually weak lines of Si II, S II, and Ca II, yet
prominent high-excitation features of Fe III. The usual Si II, S II, and 
Ca II lines develop in the post-maximum spectra, and by a few
weeks past maximum the spectra look nearly normal. However, as
Filippenko et al. (1992a) and Li 
et al. (1999) pointed out, at late times subtle differences still exist
between high-quality spectra of SN 1991T-like objects and those of normal 
SNe Ia.  SN 1999aa (Filippenko, Li, \& Leonard
1999) is similar to SN 1991T (Figure 1) in having very
weak Si II 6150 \AA\, absorption and prominent Fe III lines near
4300 \AA\, and 5000 \AA, but it also shows strong absorption centered 
at 3750 \AA\, that is probably due to Ca II H\&K, 
which is very weak in SN 1991T. 
Whether objects like SN 1999aa (to be discussed in more detail by 
Li et al. 2000c) are of an intermediate type 
between SN 1991T-like events and normal SNe Ia remains to be determined.
Not to complicate the classification scheme, however, here we still 
classify SN 1999aa-like objects as ``SN 1991T-like," even though
most SN 1991T-like objects seem to actually be like SN 1999aa. 
At this time there is widespread familiarity with SN 1991T
subclass, whereas SN 1999aa-like objects were only recently 
recognized and are unknown to most researchers. See Section
4.1 for further discussion of this issue.

The main spectral peculiarities of SN 1991bg (Filippenko et al.
1992b; Leibundgut et al. 1993; Turatto et al. 1996; Mazzali et al. 1997), 
when at maximum brightness, are the presence of a broad absorption 
trough extending from about 4100 to 4400 \AA\, due to Ti II lines, 
and  enhanced Si II 5800 \AA\, absorption relative to Si II 
6150 \AA.

SN 1986G (Phillips et al. 1987; Cristiani et al. 1992) is similar 
to SN 1991bg but less extreme. 
The Ti II 4100-4400 \AA\, absorption trough is present in the 
spectra of SN 1986G, yet is weaker than that of SN 1991bg.
Since it is difficult to differentiate SN 1986G-like objects from 
SN 1991bg-like ones without having excellent data, we have combined the two classes together,
and call them SN 1991bg-like objects hereafter, unless otherwise noted.

Various authors have shown that the spectroscopic behavior 
of SNe Ia is correlated with their photometric  behavior, which
in turn is related to their absolute magnitudes
(e.g., Nugent et al. 1995; Filippenko 1997, and references therein). SN 1991T-like 
objects rise to their maximum more slowly and decline more slowly,
and are overluminous relative to normal SNe Ia; SN 1991bg-like 
objects rise to their maximum and decline more
quickly, and are subluminous relative to normal SNe Ia. Thus light curves,
if available, can often serve as an independent check of the peculiarities 
of SNe Ia, though exceptions do exist (e.g., SN 1992bc, Maza et al. 1994; SN
1999ee, Hamuy 2000, private communication). 

\subsection{ The Sample of SNe Ia }

We have compiled a set of 45 nearby SNe Ia discovered in the sample 
galaxies of LOSS and BAOSS. This includes 38 SNe Ia discovered by
LOSS and BAOSS, and 7 SNe Ia that were initially discovered by other 
groups, but were subsequently rediscovered in the course of LOSS or BAOSS.
Table 1 lists the information for all these SNe in the following format:
Column (1): SN name. 
Column (2): host galaxy of the SN. 
Column (3): the Hubble type of the host galaxy. 
A classification followed by a ``:'' means it is uncertain. 
Column (4): the number of the International Astronomical Union (IAU)
Circular that contains the spectroscopic
classification of the SN. 
Column (5): the age of the SN at the time of the spectroscopic
classification, relative to the time of $B$-band maximum brightness. This is generally {\it not} the age of the SN at 
discovery, as it takes severals days (sometimes longer)
for the spectroscopic classification of a SN to be made.
{The quoted ages are derived either from the IAU Circular listed in 
Column (4), which are marked with a colon and generally have uncertainties
of about $\pm$5 days, or from our own light-curve database and/or 
spectral analysis, which usually have uncertainties of about $\pm$2 days.}
Column (6): discoverer of the SN. ``L'' means LOSS, ``B'' means
BAOSS, and ``O'' means other groups. 
Column (7): whether the SN is peculiar. If it is, the type of 
peculiarity is listed. Objects marked as peculiar type 
``91T$_{aa}$'' are those that are similar to 
SN 1999aa as discussed earlier. For SNe 1997cw and 1999gp, 
no spectroscopic information are available to us to determine whether they 
are SN 1999aa-like, and they are classified
as ``91T/91T$_{aa}$". 
Columns (8) to (11): evidence of the peculiarity of the SN. Specifically,
Ti II absorption (Col. 8), 
a relatively strong Si II 5800 \AA\, absorption (Col. 9),
prominent Fe III absorption lines (Col. 10), 
and weak or no Si II 6150 \AA\, absorption (Col. 11).
Our major source of information for the host galaxies of the SNe
is NED\footnotemark{}, and that for the SNe is the IAU Circulars. 

\footnotetext{The NASA/IPAC Extragalactic Database (NED) is operated by the Jet Propulsion Laboratory, California Institute
     of Technology, under contract with the National Aeronautics and Space Administration.}

One danger of using the information in the IAUCs to subclassify a SN Ia
is that it was most likely derived by the observer during, or
shortly after the observations. This ``on-the-fly'' classification might result in 
quick-look reduction errors and approximations. The subclassification also depends on the 
observer's ability to recognize the peculiarities in the spectra
of SNe Ia and the willingness to include the information in the report
to the IAUC. Sometimes  
classifications of SNe in the IAUCs
are revised because of subsequent observations or reductions.
For this reason, we count a SN Ia as peculiar only when there is direct 
evidence of peculiarity included in the classification of the SN in IAUCs, 
or we have our own spectra to support the 
classification. 
Columns (8) to (11) in Table 1 list the evidence reported in 
the relevant IAUCs. A SN is classified as
SN 1999bg/1986G-like only when there is evidence in the IAUC
that its spectrum shows Ti II absorption and a relatively strong Si II
5800 \AA\, line, while a SN 1991T-like object must show evidence
of Fe III absorption lines and weak (or absent) Si II lines. 

We have classified two SNe as peculiar even though they were not
listed as such in the IAUCs. SN 1999bh was discovered by LOSS (Li 
1999). Aldering et al. (1999) reported that the spectrum 
was of a SN Ia within one week after maximum brightness. Nugent (1999) communicated
that Aldering's spectrum of SN 1999bh resembled SN 1986G at 3 days 
past maximum. 
In addition, comparisons of its spectrum at a month past maximum to that of a normal
SN Ia (SN 1998bu; Jha et al. 1999)
and a SN 1991bg-like  SN Ia (SN 1997cn; Turatto et al. 1998) at comparable epochs (Figure 2) 
reveal that SN 1999bh is similar to SN 1997cn at this phase. 
We thus classify SN 1999bh as a SN 1986G/1991bg-like 
object.

SN 1999dg was also discovered by LOSS (Modjaz \& Li 1999). Filippenko, Metzger,
\&  Small
(1999) classified it as a SN Ia within a week past maximum brightness. A spectrum of
the SN is shown in Figure 1. Apparent Ti II absorption is seen at around 
4100-4300 \AA, and the Si II absorption at 5800 \AA\, is relatively 
strong. We classify SN 1999dg as a SN 1986G/1991bg-like object.

\section{Statistics}

\subsection{ The Observed Peculiarity Rate}

Out of the 45 SNe in our sample, 9 are SN 1991T-like and 
7 are SN 1991bg-like. The total peculiarity rate is 
36$\pm$9\%. The rate for SN 1991T-like objects is 20$\pm$7\%, 
and for SN 1991bg-like objects it is 16$\pm$6\%. The uncertainties
quoted here are derived only from Poisson statistics.
Our total peculiarity rate is substantially higher than 
the BFN result (less than 17\%); details are presented in Sec. 4.2.

\subsection{ The Intrinsic Peculiarity Rate} 

Because of the variations  among the luminosities, light-curve
shapes, and spectral evolution of different kinds of SNe Ia, there are several 
possible observational biases in the observed SN sample, 
and the observed peculiarity
rate could be either an overestimate or an underestimate of the 
intrinsic peculiarity rate. The relationship between the 
observed and the intrinsic peculiarity rates 
largely depends on the way the SNe are discovered -- that is,
on the survey methods. 

SN surveys can be generally divided into two categories, magnitude-limited
and distance-limited.
In a magnitude-limited SN survey,
the target fields are usually random regions on the sky that
contain many galaxies at different redshifts. The number of
SNe discovered depends on the detection limit of the images
(hence magnitude-limited). 
Target fields for a distance-limited SN survey, on the other hand,
are individual galaxies or clusters of galaxies. The limiting
magnitude of the survey images
may be much deeper than all the possible SNe in the sample galaxies
or clusters
(as long as they aren't heavily extinguished), 
so the number of SNe discovered largely 
depends on the number and distances of the sample galaxies or clusters
(hence distance-limited). 
A few SN surveys can be best described as a hybrid of these two 
categories. They are distance-limited, but many of the discovered
SNe are in the background of the 
target galaxies and some of them have magnitudes close to the detection
limit of the survey images. 

Li et al. (2000b; hereafter Paper I) have used Monte Carlo simulations
to study various observational biases
of SN Ia observations in SN surveys. 
They found that there are {four main observational biases.
The ``age bias," which is caused by the fact that SN 1991T-like
objects can only be easily identified with spectra taken 
prior to or near maximum brightness; the Malmquist 
bias, which is caused by the difference in luminosity among SNe Ia; the
``light-curve shape bias," which is caused by the difference in the 
light-curve shapes of SNe Ia; and the ``extinction bias," which is caused
by the fact that SN 1991T-like objects often occur in dusty, star-forming
regions and may suffer more extinction than the other objects. 
They also found }that the effect
of these biases depends on the characteristics of the SN survey,
such as distance-limited or  magnitude-limited, baseline,
and limiting magnitude. The adopted cutoff for the age bias
and the assumed extra extinction applied to SN 1991T-like objects 
for the extinction
bias also play important roles in determining the results of the 
observations. 
One result from the simulations {(Figures 11 and 15 in Paper I)} is
that for a distance-limited survey conducted in the $R$ band
with (a) the limiting distance set
to where normal SNe Ia have a peak apparent magnitude of 16.0,
(b) a limiting magnitude of 19.0, and (c) a baseline of $\le$ 10 days,
all the observational biases play negligible roles {(even
for extreme cases such as a cutoff of 1 day before maximum light for the age
bias, and an extra extinction of 0.8 mag for SN 1991T-like objects)} and 
all SNe Ia are found in the survey, yielding the intrinsic peculiarity
rate and the luminosity function of SNe Ia. 

LOSS searches about 5000 fields containing 6000 galaxies,
most of which have 
radial velocities less than 8,000 km s$^{-1}$.
BAOSS searches about 3000 fields containing some 4,000 target galaxies
(BAOSS has a larger field of view than LOSS),
most of which have radial velocities less than 6,000 km s$^{-1}$.
{For the period of this study, most galaxies in the two
searches overlapped, but whereas BAOSS searches more galaxies 
in the winter season when it has good weather, LOSS searches
more galaxies in the summer season for the same reason.}
Both surveys are condutcted in unfiltered mode, whose closest match to 
the standard passbands is $R$.
Using a Hubble constant of 65 km s$^{-1}$ Mpc$^{-1}$, and an
absolute magnitude of normal SNe Ia of $M(B)\,=\,-19.5$, we derive 
that the limiting distances for LOSS and BAOSS are where normal 
SNe Ia have apparent peak magnitudes of 16.0 and 15.3, respectively.
Figure 3 shows the number of galaxies in the LOSS sample versus
the expected peak apparent magnitude of normal SNe Ia. We  see
that more than 96\% of the expected SNe Ia in the sample galaxies
have peak apparent magnitudes brighter than 16 (the dashed vertical line).
The usual survey images of LOSS and BAOSS have a limiting magnitude of about 
19.0,  much fainter 
than the faintest normal, unextinguished SNe Ia in the samples.
The two surveys are thus limited by the distances of their sample
galaxies and belong to the distance-limited SN surveys.

Because of the high observation efficiencies 
(80 images per hour for LOSS and 60 for BAOSS), the two surveys
have short baselines. Depending on the right ascension of the galaxies,
LOSS has baselines of about 4 to 8 days and BAOSS has baselines
of 4 to 10 days. Sometimes, because of spells of poor weather,  
the intervals between observations are longer than 10 days for either
LOSS or BAOSS, but fortunately the weather in the two surveys is largely
complementary as noted above. {Since most sample
galaxies are in both surveys, the combined 
result is a baseline $\lesssim$ 10 days.} 

The Monte Carlo simulations in Paper I indicate that for
the characteristics of LOSS and BAOSS, 
essentially {\it all} SNe Ia in the LOSS and BAOSS sample galaxies should have been discovered
(Figure 9 in Paper I),
and the observed rates for the peculiar SNe Ia should equal their intrinsic
values (Figures 11 and 15 in Paper I).
[The only rare exceptions are SNe that occurred in the
host galaxy nuclei that are very bright and star-like, as in the case
of Seyfert galaxies, and SNe that occurred in galaxies that 
are well past the meridian at the beginning of the night, which 
the  two surveys don't monitor much (Filippenko et al. 2000; Li et al. 2000a).]
This means that
(1) the observed peculiarity rate 
(36\%) in our sample is close to the intrinsic one, and the $1\sigma$ uncertainty
of the rate is $\pm$9\% according to Poisson statistics; and
 (2) the observed luminosity function of SNe Ia
is intrinsic:  SN 1991bg-like objects are 16$\pm$6\%, normal
objects are 64$\pm$12\%, and SN 1991T-like objects are 20$\pm$7\% of the total
SN Ia sample. 

\section {Discussion}

\subsection{Caveats Concerning the High Peculiarity Rate}

In discussing the high peculiarity rate of SNe Ia found by
LOSS and BAOSS, it is relevant to ask whether this
galaxy sample is representative of
the general population of galaxies.
For example, the sample would be biased if the galaxies were 
selected to be predominantly spirals rather than ellipticals, since
the former have a higher SN rate than the latter. 
It is not entirely clear how such a  selection bias would affect the 
resulting peculiarity rate and luminosity function of SNe Ia,  
because the relation between the distribution of 
peculiar SNe Ia and Hubble types of the host galaxies is 
still uncertain. 
This bias, however, is only
present in the BAOSS sample during its first year of operation 
(1996, not included in the 
current study), when there 
were few elliptical galaxies in the sample. The 
situation was corrected at the beginning of 1997, when a totally
new sample was constructed without any selection bias against
particular Hubble types of  galaxies (except, perhaps, the 
Ir and dwarf galaxies, which are under-represented in all galaxy
catalogs).  A comparison of the 
Hubble-type distribution of the galaxies in the LOSS sample, 
which is constructed in the same way as the BAOSS sample but
extends to larger redshifts, with that of about 9,800
galaxies having $cz \le $ 8,000 km s$^{-1}$ in 
the catalog of de Vaucouleurs et al. (1991),  is displayed in 
Figure 4.
In the LOSS sample there seems to be relatively more 
galaxies of Hubble type Sc than other types,
but the galaxies
that are somewhat under-represented are actually the 
very late-type spirals
Scd, Sd, and Sdm. The relative fraction of elliptical galaxies 
in the LOSS sample is about 
90\%  of that in the de Vaucouleurs et al. (1991) catalog, 
so discrimination against them is not significant.

Another concern is whether the high apparent peculiarity rate of 
SNe Ia in the LOSS and BAOSS sample is partly caused by the fact that 
most of the SNe were discovered before maximum brightness, and more than
half of them were discovered  at least a week before maximum. 
At those early times, the spectrum of a SN Ia reveals the
physical conditions of the outermost parts of the ejecta,
where differences among individual SNe (e.g., metallicity) are 
most likely to reside. The high peculiarity rate found 
in our sample may thus be partly caused by the 
inhomogeneity of normal SNe Ia at early times.
In particular, we are concerned with the 
SN 1999aa-like objects. Although we classified them as 
SN 1991T-like objects in the calculations of the peculiarity
rates because their spectra show prominent high-excitation 
Fe III lines as does SN 1991T, we note the 
existence of prominent Ca II H\&K lines in their spectra,
which are weak in SN 1991T. Whether these objects are SN 1991T-like
ones, or intermediate
ones between SN 1991T-like and normal, or
normal ones caught at very early times, remains to be determined. 
One way to investigate this is by studying their photometric 
and spectroscopic behavior (Li et al. 2000c).

To make our point more clearly, Figure 5 shows the
near-maximum spectra of three SN 1999aa-like objects (SNe 1999aa,
1998es, 1999ac, all at age 1 day before maximum) and one normal object
(SN 1994D at age 3 days before maximum). 
The overall similarity among the spectra is 
striking. {\it Had the three SN 1999aa-like objects been initially 
observed at this phase or
later, they would have been classified as normal SNe Ia.} 
Unfortunately, there are no published spectra for genuine SN 1991T-like
objects at this phase (e.g., SN 1991T -- Filippenko et al. 1992a, Phillips 
et al. 1992; SN 1997br -- Li et al. 1999), so it is not clear 
whether they showed the same spectral evolution.
This comparison raises two possibilities: 

(1) The SN 1999aa-like objects may
be (photometrically) normal objects caught at very early phases.
{In other words, there may be two classes of SNe Ia that are
photometrically 
normal, but spectroscopically distinct at very early times:
those like SN 1999aa, which showed Fe III lines,
Ca II H\&K lines and no Si II 6150 \AA\, 
line; and those like SN 1994D (e.g., Patat et al. 1996; Filippenko 1997)
and SN 1990N (Leibundgut et al. 1991), which showed a strong 
Si II 6150 \AA\, line and looked normal.} 
If the SN 1999aa-like objects (including uncertain objects such
as SN 1997cw and 
SN 1999gp) are counted as normal SNe Ia, 
the peculiarity rate in the LOSS and BAOSS sample becomes only 18$\pm$4\%
instead of 36$\pm$9\%,  in agreement with 
the rate reported by BFN (less than 17\%).

(2) The cutoff of the age bias could be
as early as  1 day before peak brightness if the SN 1999aa-like
objects indeed belong to the class of SN 1991T-like objects.
{If this is true, the simulations in Paper I indicate
that LOSS and BAOSS should still discover all the SNe Ia 
early enough and yield the intrinsic peculiarity rate and luminosity 
function (Figure 15 in Paper I). Inspection of Table 1, however, shows 
that this may not be the case. 21 out of the 45 SNe Ia in
our sample were classified later than one day before maximum, 
so the age bias would be significant if the cutoff were
one day before maximum and our rate of SN 1991T-like objects
would be only a lower limit. There are two possible reasons for this
discrepency between the observations and the simulations: (i) the 
age estimates in Table 1 are quite uncertain, especially
those with ``0:" which were reported in the IAU Circulars as 
``near maximum";  and (ii) the gap between the time of discovery and
the time of spectroscopic classification is simulated as a 
random number between 2 and 5 days in Paper I, whereas in
real observations,
the gap could be much larger. For example, 6 out of the 45 SNe Ia
in Table 1 were spectroscopically classified more than 10 days
after their discovery;  the observations will therefore 
have a more serious age bias than the simulations done
in Paper I.

An earlier cutoff for the age bias also has a direct impact
on explaining the null discovery of SN 1991T-like objects in the 
Cal\'an/Tololo SN survey and the SN surveys done at high redshifts
(see discussions below). }

The third concern is that the LOSS and BAOSS sample
may suffer from uncertainties caused by small-number 
statistics -- there are only 45 SNe Ia.
While this is not a much smaller sample than the 84 SNe Ia considered
by BFN, a more accurate peculiarity rate and a more
precise luminosity function for SNe Ia will be obtained when
more SNe are discovered by the two surveys.

{The high intrinsic rate of SN 1986G/1991bg-like and SN 1991T/1999aa-like
objects also suggests that they are not uncommon among
observed SNe Ia, 
and raises the question of whether it is appropriate to 
categorize them as ``peculiar SNe Ia." We call
them ``peculiar" for historical reasons -- they were  rare until the past two
years, and they show distinct features in their spectra. Moreover,
in comparing our results with those of BFN, it is appropriate to 
adopt the same terminology. With such a high
peculiarity rate, however, these events become an important part of
the whole SN Ia sample, and they may be part of a spectroscopic/photometric
sequence of SNe Ia. For example, the SN 1999aa-like objects may be 
the ``missing link" between SN 1991T-like objects and normal events.
More spectroscopic and photometric observations are needed to 
test these suggestions.} 

\subsection{Comparison with the BFN Results}

An advantage of the LOSS and BAOSS sample is that the SNe Ia
came from well-defined SN surveys, which enables 
the observational biases to be well studied. In addition, 
the assumed extinction of
SN 1991T-like objects and the adopted cutoff of the age 
bias, which are at present poorly constrained by observations, play  
negligible roles in determining the observed peculiarity 
rate (Paper I) -- for example, all SNe are discovered regardless of the 
extinction used for SN 1991T-like objects (within reasonable limits).
Even if SN 1991T-like
objects had the apparent brightness of SN 1991bg-like objects
(which means an extra $R$-band extinction of $A$=1.4 mag), all of them 
would still be discovered in the LOSS and BAOSS surveys
(just as all SN 1991bg-like objects are discovered).

The situation for the SN sample used by BFN, however, is
very different. The SNe were discovered 
by many different groups using various survey methods.
A large fraction of the SNe 
were discovered before 1985 when there were no
systematic distance-limited SN surveys and few 
opportunities for rapid spectral classification. 
Most of the SNe in the BFN sample were discovered after 
maximum brightness and the spectra used to classify them 
were obtained even later. It is thus 
very difficult to disentangle the various observational
biases in the sample, {many of which were indeed mentioned
by BFN.} Moreover, since many of BFN's SNe came
from magnitude-limited surveys, the uncertain extinction
of SN 1991T-like objects makes it difficult to derive 
the intrinsic peculiarity rate from the observed one.
The observed peculiarity rate as derived by BFN (less than
17\%) is thus a very rough estimate of the intrinsic 
one. The relatively small number for the peculiarity rate 
quoted by BFN can be understood primarily in the following way:
(1) the sample underestimates the rate of SN 1991bg-like
objects because of the light-curve shape bias and the Malmquist bias;
and
(2) the sample underestimates the rate of SN 1991T-like
objects because of the age bias. 
These two reasons also probably explain why SN 1991T-like and 
SN 1991bg-like objects were not recognized
before 1990.

\subsection {Comparison with the High-Redshift Results}

During the past few years, two groups have presented strong evidence that
the expansion of the Universe is accelerating rather than decelerating
(Riess et al. 1998; Perlmutter et al. 1999). This surprising result comes
from distance measurements to about fifty SNe Ia in the redshift 
range $z$ = 0.1 to 1, and is based on the assumption that there are no significant
differences between the SNe Ia at high redshift and their low-redshift
counterparts.  Although this assumption is supported to first order  by comparisons 
of the photometric and spectroscopic properties of SNe Ia at high
and low redshifts (e.g., Riess et al. 1998, 2000; Coil et al. 2000), there is also some evidence
for differences between SNe Ia at different redshifts. In particular,
Riess et al. (1999c) tentatively point out that the risetime (from
explosion to maximum brightness) of low-redshift
SNe Ia as determined by Riess et al. (1999b)
differs from that of their high-redshift
counterparts as determined by Groom (1998) and 
Goldhaber (1998), although a new measurement of the 
high-redshift risetime by Aldering, Knop, \& Nugent (2000) 
diminishes the significance of this difference. 
Also, we note that Howell, Wang, \& Wheeler (2000) point out
a difference in the low-redshift and high-redshift radial distributions
of SNe Ia in their host galaxies. There may additionally be systematic differences
in the colors of low-redshift and high-redshift SNe Ia (Falco et al. 1999).

The peculiarity rate of SNe Ia at different redshifts, if taken at 
face value, may be another preliminary
indication that the high-redshift SNe Ia differ to some extent from
their low-redshift counterparts. Our study shows that the 
peculiarity rate for the nearby SNe Ia is $\sim$ 36\%.
For the high-redshift SNe Ia, however, there has not been  a single peculiar
SN Ia reported among the $\sim$ 50 spectroscopically classified SNe (Riess et al. 1998;
Perlmutter et al. 1999; Adam Riess 1999, private communication; 
Peter Nugent 1999, private communication). The high-redshift SN searches 
are magnitude-limited surveys conducted in approximately the rest-frame $B$ band
with a baseline of $\sim$ 20 days in the rest frame. With
the intrinsic luminosity function for SNe Ia derived in Section 3, 
the rates of SN 1991bg-like and 
SN 1991T-like objects can be correctly derived for the magnitude-limited
surveys using the Monte Carlo simulations in Paper I. 
The results are shown in Figure 6 for {six cases, with
SN 1991T having different amounts of adopted extra extinction in the $B$ band   
and two cutoffs (7 days after and 1 day before peak brightness) for
the age bias}. We see that the rate for SN 1991bg-like objects is less than
2\%, regardless of the adopted extra extinction for SN 1991T-like objects and
the cutoff of the age bias,
so it is not a surprise that there have been few or no such objects
discovered at high redshifts. However, {if a cutoff of +7 days is used to 
simulate the age bias (the left panels in Figure 6),} the rate for SN 1991T-like
objects should be  27.3\%, 18.6\%, and 12.0\% for the 
cases when SN 1991T-like objects have no extra extinction,  0.4 mag of extra
extinction, and 0.8 mag of extra extinction, 
which means that there should be about 14$\pm$3, 9$\pm$3, and 6$\pm$2 SN 1991T-like
objects among the 50 high-redshift SNe Ia, respectively.  The
fact that there may be no clear
SN 1991T-like objects in the high-redshift sample is 
thus surprising. A natural explanation is that there may be evolution in
the characteristics of SN Ia explosions at different redshifts.
The disappearance of SN 1991T-like objects at high 
redshift may result from the loss of certain
progenitor channels at high redshift due to a possibly 
redshift-dependent variation in the mass, composition, and metallicity of SN 
Ia progenitors (e.g., Ruiz-Lapuente \& Canal 1998; Livio 2000). 

On the other hand, there are several other possible
explanations for the null discovery of
SN 1991T-like objects at high redshift. These include the following.

(1) The SNe Ia at high redshifts are very faint, and spectroscopy of them
is challenging even with the world's largest optical telescopes.
Given the fact that it is often not possible even to 
classify SNe at high redshift, it is far more difficult to detect peculiarities
in their spectra. This problem is exacerbated by the inability to 
view the region around  redshifted Si II 6150 \AA\, in most of the objects; it 
is at wavelengths inaccessible to optical spectrgraphs, and the 
near-infrared sky is bright. This leaves mainly the Ca II
H\&K lines as discriminants of peculiarity, and makes it difficult to 
rule out SN 1999aa-like objects (which, recall, we include here in 
the SN 1991T subclass). Since most of the spectra of high-redshift
SNe Ia have not yet been published, we cannot assess whether the 
absence of reported peculiarity is significant. 

(2) The SN 1991T-like objects may be extinguished by more than the maximum
extinction of $A$=0.8 mag displayed in Figure 6. This
higher extinction
can easily make the apparent difference
in the peculiarity rate at low and high redshifts statistically
insignificant. To investigate this possibility we have repeated the
Monte Carlo
simulations in Paper I for a wide range of assumed
extinctions for the SN 1991T-like objects,  
 and the results are shown in Figure 7. As expected,  the observed rate of SN 1991T-like
objects drops dramatically with an increasing adopted extinction. However,
to make the null discovery of SN 1991T-like objects among the 50 high-redshift
SNe Ia (assuming all have good spectra) be within the 3$\sigma$, 2$\sigma$, and 1$\sigma$
confidence levels of the nearby rate, extinctions of 
$A$=0.9, 1.4, and 2.1 mag must be adopted for the SN 1991T-like
objects.  Although four known SN 1991T-like objects
showed a significant amount of reddening $E(B-V)$ [SN 1991T has 0.13 mag
(Filippenko et al. 1992a),
SN 1995ac has 0.17 mag (Riess et al. 1999a), SN 1995bd has 0.5 mag
 (Riess et al. 1999a), and SN 1997br
has 0.35 mag (Li et al. 1999)], the intrinsic reddening due to the 
SN environment is only $E(B-V) = 0.11$ mag for SN 1991T,  0.13 for
SN 1995ac, 0.00 for SN 1995bd, and  0.24 for 
SN 1997br, when the Galactic component of the extinction is subtracted
according to the map of Schlegel, Finkbeiner, \& Davis (1998). Thus, the
average intrinsic reddening for the four known SN 1991T-like objects 
is only $E(B-V) = 0.12$ mag. Assuming a normal extinction curve,  the extinction
in the $B$ band would be about 0.5 mag. The requisite large extinction 
($A\ge$1.1 mag) to account for the difference between the observed rate of SN 1991T-like objects
at different redshifts therefore might not be plausible.

{(3) There may be a more serious age bias -- that is, a cutoff earlier
than +7 days for 
the age bias. In the discussions of Figure 5 we pointed out that
the cutoff for the age bias could be as early as 1 day before 
peak brightness. The results for such an early cutoff are shown in the 
right panels of Figure 6.
For a baseline of 20 days, the rate of SN 1991T-like objects
is only 10\%, 6\%, and 4\% for the case when SN 1991T-like objects 
have 0, 0.4 mag, and 0.8 mag of extra extinction, 
which means that there should be about 5$\pm$2, 3$\pm$2, and 2$\pm$1
SN 1991T-like objects among the 50 high-redshift SNe Ia, respectively.
The results with an early cutoff for the age bias indicate that the null discovery of SN 1991T-like objects in the 
high-redshift SN Ia sample is not significantly different 
from the cases when SN 1991T-like objects have extra extinction
($A$=0.4 or 0.8 mag).}

The results from the Cal\'an/Tololo SN Survey (CTSS) provide
another compelling reason to believe that the difference between 
the peculiarity rate of SNe Ia at high and low redshifts is caused
by the above three reasons rather than by the evolution of SNe Ia at
different redshifts.
The SNe Ia discovered in CTSS
have redshifts of about 0.05, which is 
intermediate to those found at high redshift
($z \approx $ 0.5) and those found by LOSS and BAOSS ($z < 0.03$).
If the difference in the peculiarity rate of SNe Ia at high
and low redshifts were caused by the evolution of SNe Ia, 
we would expect the rate of SN 1991T-like objects in 
CTSS  to be in between these two 
rates.  But the fact is that there are {\it no } apparent SN 1991T-like 
objects among the 31 SNe Ia discovered in the course of CTSS.
CTSS and the 
high-redshift SN surveys have very similar characteristics:
all of them are magnitude limited
with a baseline of about 20 days, and all of them are done in
the rest-frame $B$ band. The null discovery of SN 1991T-like 
objects in these surveys at different redshifts strongly implicates
the survey characteristics rather than the evolution of SNe Ia. 

{There is also evidence from the CTSS data that
indicates serious observational biases and further supports
this explanation. 
Mark Phillips (2000, private communication) reports that only
23\% of the SNe Ia discovered in CTSS
have spectra at maximum or earlier, indicating
the existence of a serious age bias. In contrast,
78\% of the SNe in Table 1 were observed spectroscopically at or 
before maximum. Just on the basis of these numbers, one
might expect that 2 of every 3 SN 1991T-like events
would be classified as normal in CTSS. 
Hence, for the 31 SNe discovered, only 1.8$\pm$1 SN 1991T-like
objects would have been expected if the intrinsic rate of 
SN 1991T-like events were 20\%. This small number 
is not significantly inconsistent with the fact that none
of these events were found.}

\subsection{The Progenitor Systems of SNe Ia}

As discussed in Section 1, the exact nature of the progenitor systems
of SNe Ia is still unknown. Until this problem is solved, one 
cannot be fully confident in the use of SNe Ia for cosmological 
distance determinations. 
The favored models (for reviews see Branch et al. 1995, and Livio 2000)
include the double-degenerate models which involve the coalescence of
C-O white dwarf pairs, and the single-degenerate  models which
involve a single white dwarf accreting material from a subgiant
or giant companion (systems like supersoft X-ray sources, and symbiotics).

So far the homogeneity of SNe Ia has provided a very strong
restriction to the models. Because it was thought that nearly 90\% of all
SNe Ia form a homogeneous class in terms of their
spectra, light curves, and luminosities (the BFN
results), researchers have been working hard to choose a single
model over all others. It has been proposed that SNe Ia in late-type and
early-type galaxies may have different progenitors (e.g., Della Valle
\& Livio 1994; Ruiz-Lapuente, Burkert, \& Canal 1995), but 
the homogeneity constraints of SNe Ia have served as an argument against
this.

The high peculiarity rate found in our study suggests that 
homogeneity is no longer such a strong constraint for the 
progenitor systems of SNe Ia. The high peculiarity rate actually 
may favor the existence of different progenitor systems; with 
more than one class of progenitor system, SNe Ia are more likely to have 
different types of photometric and spectroscopic properties. For
example, perhaps SN 1991bg-like objects come from double-degenerate progenitor systems,
while SN 1991T-like objects have single-degenerate progenitors, 
or vice versa. A full discussion,
however, is beyond the scope of this paper.

\section {Conclusions}

We have compiled a set of 45 SNe Ia discovered
in the LOSS and BAOSS galaxy samples (up through SN 2000A).
We find that
the total observed peculiarity rate is 36$\pm$9\%; the rates are
16$\pm$7\% and 20$\pm$7\% for 
SN 1991bg-like and SN 1991T-like objects, respectively.
Only 64$\pm$12\% of observed SNe Ia are normal.
Our peculiarity
rate is substantially higher than that reported by Branch, 
Fisher, \& Nugent (1993).

LOSS and BAOSS are distance-limited SN surveys with a limiting
distance where normal SNe Ia have a peak apparent magnitude of 
about 16.0 and 15.3, respectively. The limiting magnitude of the
two surveys is about 19.0 and the baseline is $\le$
 10 days. Monte Carlo simulations done by Li et al. (2000b)
indicate that essentially all SNe Ia should have been found
in the LOSS and BAOSS galaxies. This implies that the peculiarity
rate in our sample is very close to the intrinsic one, within the 
uncertainties of small-number statistics. Moreover,  the luminosity
function of SNe Ia in our sample is also intrinsic.

We have discussed some other selection effects that may 
affect our results. In particular, we find that our galaxy 
samples are representative of the general population of bright
galaxies, and that the
SN 1999aa-like objects (which are similar in some ways
to SN 1991T and are included here in the SN 1991T subclass) 
play an important role in 
determining the peculiarity rates.

The high rate of peculiar low-redshift SNe Ia is very different
from the preliminary result found at high redshift, which is 
zero.
However, it may be adequate
to explain the difference in terms of insufficient observation quality
at high redshift, 
extinction, and/or a serious age bias for SN 1991T-like
objects. If so,  it is not necessary to invoke the
evolution of SNe Ia with redshift.

The high peculiarity rate at low redshifts also suggests that homogeneity is 
no longer such a strong argument against the existence of different
progenitor systems for SNe Ia.

\acknowledgements{
We acknowledge the referee, Mark Phillips, for his very useful comments
on the manuscript. We also thank M. Hamuy, D. C. Leonard,  and T. Matheson
for helpful discussions. The work of A. V. F.'s group at U.C. Berkeley 
is supported by National Science Foundation 
grants AST-9417213 and AST-9987438, as well as by NASA grant AR-08006 from the Space
Telescope Science Institute, which is operated by AURA, Inc.,
under NASA contract NAS5-26555.
 KAIT was made possible by generous donations from Sun Microsystems Inc.,
 the Hewlett-Packard Company, AutoScope Corporation, Lick Observatory,   
 the National Science Foundation, the University of 
 California, and the Sylvia and
 Jim Katzman Foundation.
We are also grateful to the National Science Foundation of China for their
support of the Beijing Astronomical Observatory. 
}

\newpage

\begin{figure}
\caption{Spectra of SNe Ia prior to or near maximum brightness. 
Spectra of the normal SN Ia 1994D, two SN 1991T-like
objects (SNe 1991T and 1999aa), and three SN 1991bg-like 
objects (SNe 1999da, 1997cn, and 1999dg) are shown.
All spectra have been deredshifted.
Notice the strong, broad Ca II H\&K lines in the spectrum of
SN 1999aa;  they are very weak in SN 1991T.}
\label{1}
\end{figure}

\begin{figure}
\caption{Comparison of a spectrum of SN 1999bh with that of 
the normal SN Ia 1998bu and that of a SN 1991bg-like object
SN 1997cn, all obtained about a month past maximum. All spectra
have been deredshifted.}
\label{2}
\end{figure}

\begin{figure}
\caption{The number of galaxies in the LOSS galaxy sample versus
the expected peak apparent magnitude of SNe Ia in the galaxies. 
}
\label{3}
\end{figure}

\begin{figure}
\caption{The ratio of the LOSS sample galaxies to the RC3 galaxies
for different galaxy Hubble types. A total of 9,800 galaxies is used from
RC3. The ratio has been normalized by the different
number of galaxies in the two sets of galaxies. The dotted line is
ratio = 1. }
\label{4}
\end{figure}

\begin{figure}
\caption{Spectra of SNe near maximum brightness. Spectra of three
SN 1991T-like (actually 1999aa-like) objects are shown with that of the normal
SN Ia 1994D.}
\label{5}
\end{figure}

\begin{figure}
\caption{The observed peculiarity rate for the magnitude-limited surveys in the $B$ band.
Here the intrinsic peculiarity rate of SNe Ia found in the LOSS and BAOSS
is adopted. The left panels show the results for a cutoff of +7 days
for the age bias, while the right panels show those for a cutoff
of $-$1 day. Three cases with SN 1991T-like objects having different
extinctions are shown for each cutoff. In each case,
the rate of SN 1991bg-like objects is shown as a dashed line, the
SN 1991T-like objects as a dash-dotted line, and the total peculiarity
rate as a solid line. See Paper I for the details of the 
Monte Carlo simulations.}
\label{6}
\end{figure}

\begin{figure}
\caption{The observed rate of SN 1991T-like objects in magnitude-limited 
surveys in the $B$ band with a baseline of 20 days and different
adopted extinctions. As in Fig. 6, the intrinsic rate found in
LOSS and BAOSS is adopted. }
\label{7}
\end{figure}


\newpage
\begin{thebibliography}{}

\bibitem{}Aldering, G., Knop, R., \& Nugent, P. 2000, AJ, 119, 2110
\bibitem{}Aldering, G., et al. 1999, IAU Circ. 7138
\bibitem{}Barbon, R., et al. 1990, A\&A, 237, 79
\bibitem{}Branch, D., Fisher, A., \& Nugent, P. 1993, AJ, 106, 2383 (BFN)
\bibitem{}Branch, D., \& Tammann, G. A. 1992, ARAA, 30, 359
\bibitem{}Branch, D., et al. 1983, ApJ, 270, 123
\bibitem{}Branch, D., et al. 1995, PASP, 107, 1019
\bibitem{}Coil, A. L., et al. 2000, ApJ, submitted
\bibitem{}Cristiani, S., et al. 1992, A\&A, 259, 63
\bibitem{}Della Valle, M., \& Livio, M. 1994, ApJ, 423, L31
\bibitem{}de Vaucouleurs, G. 1991, {Third Reference Catalogue of Bright Galaxies} (Springer-Verlag, New York) (RC3)
\bibitem{} Falco, E., et al. 1999, ApJ, 523, 617 
\bibitem{}Filippenko, A. V. 1997, ARAA, 35, 309
\bibitem{}Filippenko, A. V., Li, W. D., \& Leonard, D. C., 1999, IAU Circ.  7108
\bibitem{}Filippenko, A. V., Metzger, M. R., \& Small, T. A., 1999, IAU Circ. 7239
\bibitem{}Filippenko, A. V., et al. 1992a, ApJ, 384, L15
\bibitem{}Filippenko, A. V., et al. 1992b, AJ, 104, 1543
\bibitem{}Filippenko, A. V., et al. 2000, in preparation
\bibitem{}Goldhaber, G. 1998, BAAS, 30, 1325
\bibitem{}Groom, D. E. 1998, BAAS, 30, 1419
\bibitem{}Hamuy, M., et al. 1996, AJ, 112, 2398
\bibitem{}Harkness, R. P., \& Wheeler, J. C. 1990, in {Supernovae}, ed. A. G. Petschek (New York: Springer-Verlag), 1
\bibitem{}Howell, D. A., Wang, L. F., \& Wheeler, J. C., 2000, ApJ, 530, 166
\bibitem{}Jeffery, D. J., et al.  1992, ApJ, 397, 304
\bibitem{}Jha, S., et al. 1999, ApJS, 125, 73
\bibitem{}Kirshner, R. P., et al. 1973, ApJ, 185, 303
\bibitem{}Kirshner, R. P., Oke, J. B., Penston, M. V., \& Searle, L. 1973, ApJ, 185, 303
\bibitem{}Kirshner, R. P., et al. 1993, ApJ, 415, 589
\bibitem{}Leibundgut, B., et al. 1991, ApJ, 371, L23 
\bibitem{}Leibundgut, B., et al. 1992, ApJ, 401, L49 
\bibitem{}Leibundgut, B., et al. 1993, AJ, 105, 301
\bibitem{}Li, W. D. 1999, IAU Circ. 7135
\bibitem{}Li, W. D., et al. 1996, IAU Circ. 6379
\bibitem{}Li, W. D., et al. 1999, AJ, 117, 2709
\bibitem{}Li, W. D., et al. 2000a, in {Cosmic Explosions}, eds. S. S. Holt and 
W. W. Zhang (New York: American Institute of Physics), 103
\bibitem{}Li, W. D., et al. 2000b, ApJ, submitted
\bibitem{}Li, W. D., et al. 2000c, in preparation
\bibitem{}Livio, M. 2000, in {Type Ia Supernovae: Theory and Cosmology} 
(Cambridge: Cambridge Univ. Press), in press
\bibitem{}Livio, M. 2000, astro-ph/9903264
\bibitem{}Maza, J., et al. 1994, ApJ, 424, L107
\bibitem{}Mazzali, P. A., et al. 1995, A\&A, 297, 509
\bibitem{}Mazzali, P. A., et al. 1997, MNRAS, 284, 151
\bibitem{}Modjaz, M., \&  Li, W. D., 1999, IAU Circ. 7229
\bibitem{}Nugent, P., et al. 1995, ApJ, 455, L147
\bibitem{}Patat, F., et al. 1996, MNRAS, 278, 111
\bibitem{}Patat, F., et al. 1997, A\&A, 317, 423
\bibitem{}Perlmutter, S., et al. 1997, ApJ, 483, 565
\bibitem{}Perlmutter, S., et al. 1999, ApJ, 517, 565
\bibitem{}Phillips, M. M., 1993, ApJ, 413, L105
\bibitem{}Phillips, M. M., et al. 1987, PASP, 99, 592
\bibitem{}Phillips, M. M., et al. 1992, AJ, 103, 1632
\bibitem{}Phillips, M. M., et al. 1999, AJ, 118, 1766
\bibitem{}Reiss, D. J., et al. 1998, AJ, 115, 26
\bibitem{}Riess, A. G., Press, W. H., \& Kirshner, R. P. 1996, ApJ, 473, 588 
\bibitem{}Riess, A. G., et al. 1998, AJ, 116, 1009
\bibitem{}Riess, A. G., et al. 1999a, AJ, 117, 707
\bibitem{}Riess, A. G., et al. 1999b, AJ, 118, 2675
\bibitem{}Riess, A. G., et al. 1999c, AJ, 118, 2668 
\bibitem{}Riess, A. G., et al. 2000, ApJ, 536, 62
\bibitem{}Ruiz-Lapuente, P., et al. 1992, ApJ, 387, L33
\bibitem{}Ruiz-Lapuente, P., Burkert, A., \& Canal, R. 1995, ApJ, 447, L69
\bibitem{}Ruiz-Lapuente, P., \& Canal, R. 1998, ApJ, 497, L57
\bibitem{}Schlegel, D. J., Finkbeiner, D. P., \& Davis, M. 1998, ApJ, 500, 525
\bibitem{}Schmidt, B. P., et al. 1998, ApJ, 507, 46
\bibitem{}Treffers, R. R., et al. 1997, IAU Circ. 6627
\bibitem{}Turatto, M., et al. 1996, MNRAS, 283, 1
\bibitem{}Turatto, M., et al. 1998, AJ, 116, 2431
\bibitem{}Wells, L. A., et al. 1994, AJ, 108, 2233

\end{thebibliography}
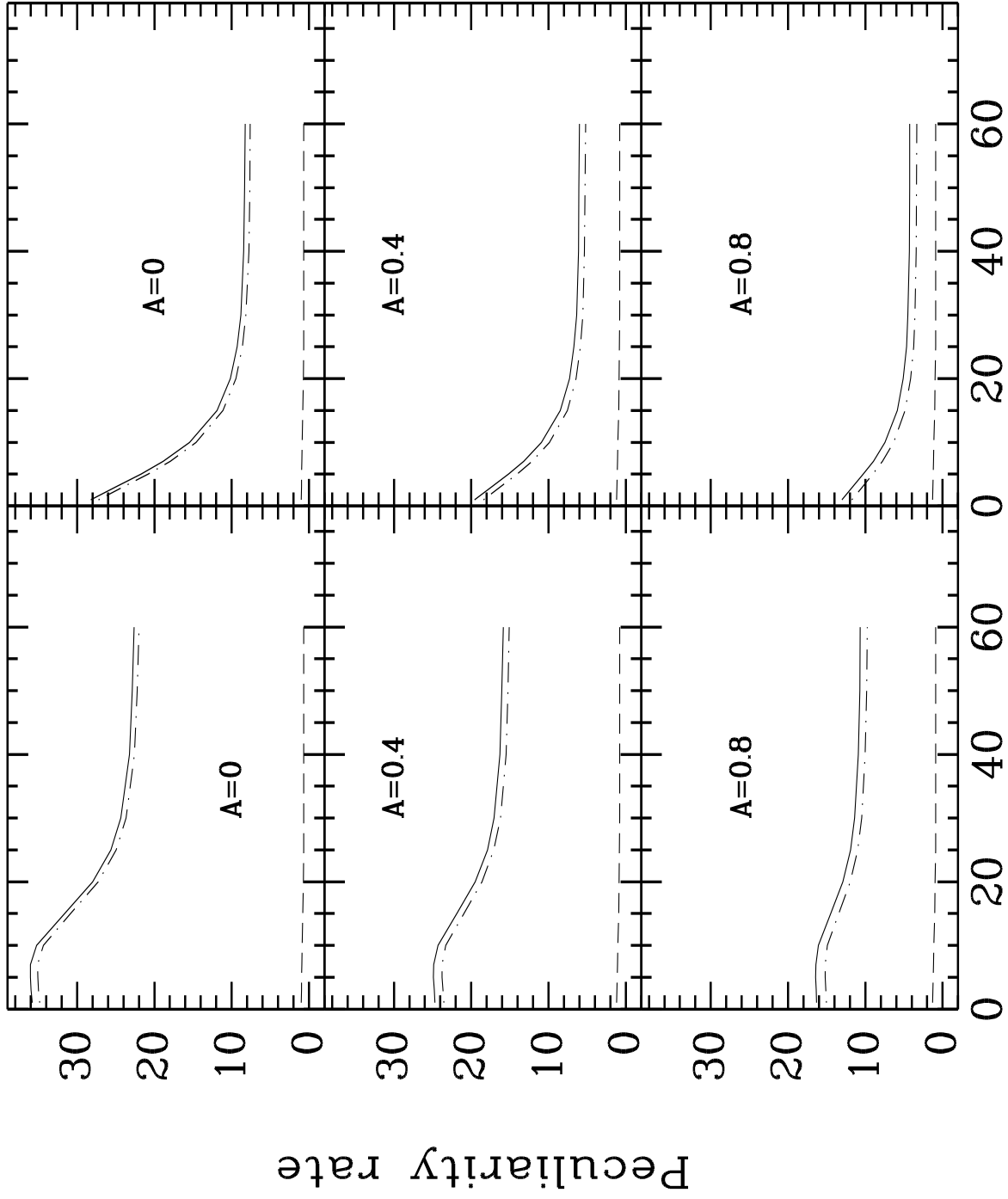
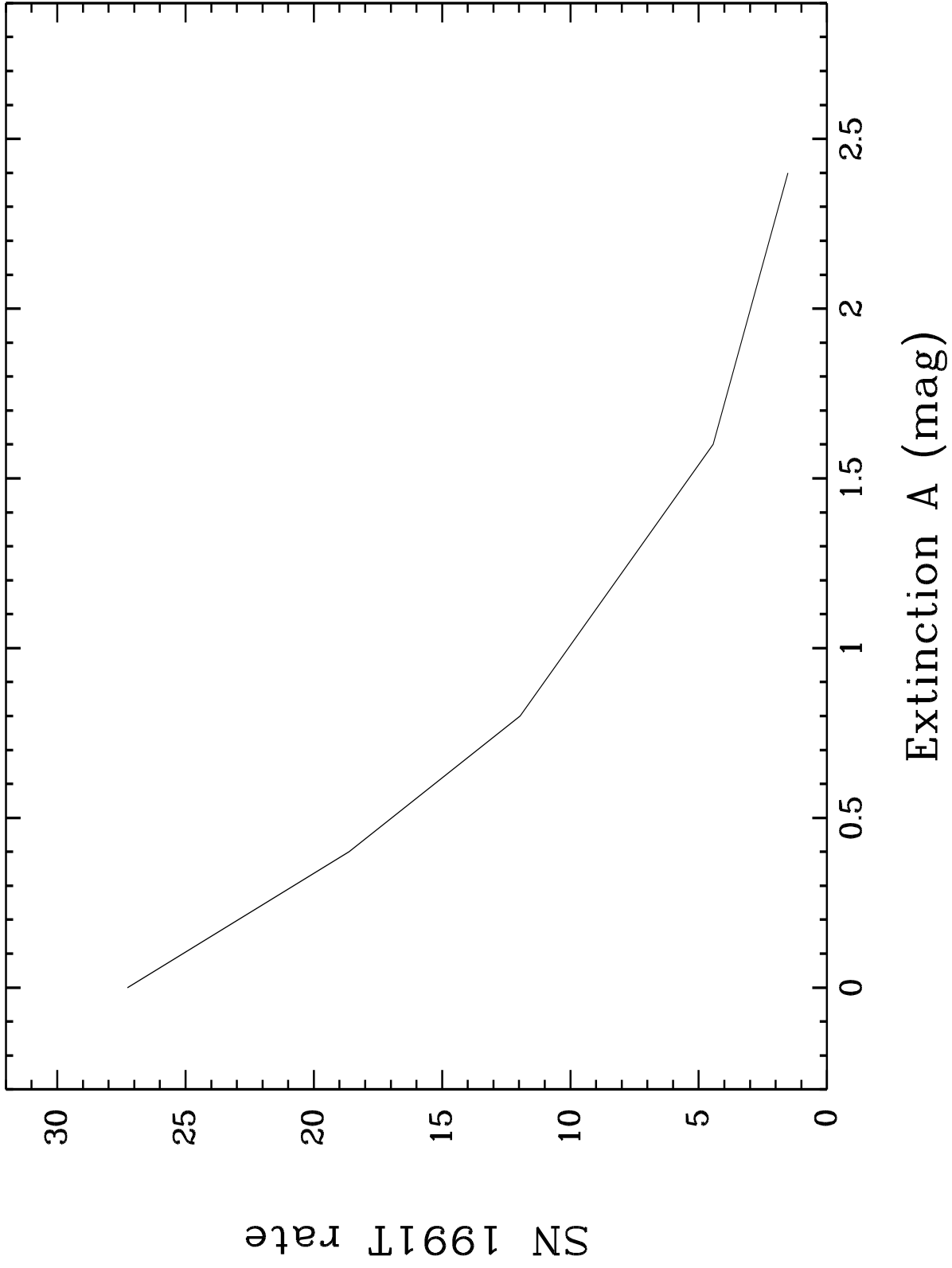
\end{document}